\documentclass{ifacconf}

\usepackage{graphicx}      %
\usepackage{natbib}        %

\usepackage{amsmath,amssymb,amsfonts}
\usepackage{mathtools}
\usepackage{xcolor}
\usepackage{enumerate}
\usepackage{lipsum}
\usepackage{accents}
\usepackage{xspace}
\usepackage{multicol, multirow}

\makeatletter
\let\old@ssect\@ssect %
\makeatother

\usepackage{natbib}
\usepackage[hidelinks]{hyperref}
\hypersetup{
    colorlinks,
    linkcolor={blue!50!black},
    citecolor={black},
    urlcolor={blue!50!black}
}

\makeatletter
\def\@ssect#1#2#3#4#5#6{%
  \NR@gettitle{#6}%
  \old@ssect{#1}{#2}{#3}{#4}{#5}{#6}%
}
\makeatother

\newcommand{\smallconc}[2]{\begin{bsmallmatrix} #1 \\ #2 \end{bsmallmatrix}}
\newcommand{\conc}[2]{\begin{bmatrix} #1 \\ #2 \end{bmatrix}}

\newcommand{\leqse}{\leq_{\mathrm{SE}}}

\newcommand{\Tgeq}{\calT_{\geq0}}
\newcommand{\ie}{\emph{i.e.}}

\newcommand{\overcirc}[1]{\mathring{#1}}

\newcommand{\R}{\mathbb{R}}

\newcommand{\IR}{\mathbb{IR}}

\newcommand{\calT}{\mathcal{T}}

\newcommand{\sfE}{\mathsf{E}}
\newcommand{\sfF}{\mathsf{F}}

\newcommand{\sfJ}{\mathsf{J}}

\newcommand{\sfM}{\mathsf{M}}

\newcommand{\ul}[1]{\underline{#1}}
\newcommand{\ula}{\ul{a}}
\newcommand{\ulb}{\ul{b}}

\newcommand{\ulw}{\ul{w}}
\newcommand{\ulx}{\ul{x}}
\newcommand{\uly}{\ul{y}}

\newcommand{\ulA}{\ul{A}}
\newcommand{\ulB}{\ul{B}}

\newcommand{\ol}[1]{\overline{#1}}
\newcommand{\ola}{\ol{a}}
\newcommand{\olb}{\ol{b}}

\newcommand{\olw}{\ol{w}}
\newcommand{\olx}{\ol{x}}
\newcommand{\oly}{\ol{y}}

\newcommand{\olA}{\ol{A}}
\newcommand{\olB}{\ol{B}}

\newcommand{\immrax}{\texttt{immrax}\xspace}

\begin{document}
\begin{frontmatter}

\title{\immrax: A Parallelizable and Differentiable Toolbox for Interval Analysis and Mixed Monotone Reachability in JAX \thanksref{footnoteinfo}} 

\thanks[footnoteinfo]{This work was supported in part by the Air Force Office of Scientific Research under award FA9550-23-1-0303 and the National Science Foundation under award \#2219755.}

\author[First]{Akash Harapanahalli} 
\author[Second]{Saber Jafarpour} 
\author[First]{Samuel Coogan}

\address[First]{School of Electrical and Computer Engineering, Georgia Institute of Technology, GA, USA 30318, \url{{aharapan,sam.coogan}@gatech.edu}}
\address[Second]{Department of Electrical, Computer, and Energy Engineering, University of Colorado, Boulder, CO, USA 80309, \url{saber.jafarpour@colorado.edu}.}

\begin{abstract}                %
We present an implementation of interval analysis and mixed monotone interval reachability analysis as function transforms in Python, fully composable with the computational framework JAX. The resulting toolbox inherits several key features from JAX, including computational efficiency through Just-In-Time Compilation, GPU acceleration for quick parallelized computations, and Automatic Differentiability. We demonstrate the toolbox's performance on several case studies, including a reachability problem on a vehicle model controlled by a neural network, and a robust closed-loop optimal control problem for a swinging pendulum.
\end{abstract}

\end{frontmatter}

\section{Introduction}

Interval analysis is a classical field concerned with bounding the output of mappings across uncertain inputs~\citep{LJ-MK-OD-EW:01}. For dynamical systems, interval analysis provides a computationally cheap, scalable, and sound approach for studying the effects of uncertainty, through differential inequalities~\citep{KS-JKS:17} and mixed monotone embeddings~\citep{SJ-AH-SC:23c}. 
While these methods have been studied extensively in the literature, most implementations either fail to (i) utilize the key computational breakthroughs in the last decade in parallel processing on GPUs and TPUs; and/or (ii) address the challenges of the modern learning-enabled control system, such as efficient gradient computation. 
JAX~\citep{jax2018github} is an evolving numerical computation framework for Python developed by researchers at Google Deepmind. At its core, JAX is a framework for \emph{composable function transformations}, \emph{i.e.}, transforms that take functions as input and return new functions with some desired property. 
For instance, \verb|jit| uses XLA to transform a function into a compiled program executed on CPU/GPU/TPU, \verb|grad|/\verb|jacfwd|/\verb|jacrev| return a new function evaluating the input function's derivative using either reverse- or forward-mode autodifferentiation from Autograd (\emph{autodiff}), and \verb|vmap| transforms a function into a new version that parallelizes its execution over several different inputs (\emph{vectorization}). 
These transformations can be composed an arbitrary number of times.
In this paper, we use these features to create an efficient, differentiable framework for interval analysis and interval reachability.

\subsubsection{Literature Review}
There are several existing tools for interval analysis and interval reachability analysis. 
To our knowledge, most of these tools do not support GPU parallelization and/or computation of gradients.
CORA is a MATLAB toolbox with interval and polytopic arithmetic capabilities~\citep{CORA}, JuliaReach is a Julia toolbox supporting interval analysis and taylor model abstractions~\citep{SB-etal:19}.
In previous work, we developed an interval analysis extension for \texttt{numpy} called \texttt{npinterval}~\citep{AH-SJ-SC:23b}, and a mixed monotone interval reachability~\citep{SC-MA:15b} tool called called \texttt{ReachMM}~\citep{SJ-AH-SC:23c}.
However, these lack support beyond basic CPU capabilitites.
For pararallization, there have been some recent developments in the reachability literature including ReachNN*~\citep{JF-CH-XC-WL-ZQ:20} which is GPU accelerated
and POLAR-Express~\citep{YW-etal:23} which supports CPU threading, but not GPU processing. 

\verb|jax_verify| is a Python library that can compute the natural inclusion function from Proposition~\ref{prop:Natural}---however, it is restricted to the class of functions used for neural networks.
The user interface is not compositional, and the usage with existing JAX transformations such as \verb|jit| and \verb|vmap| requires the user to manually convert the \verb|IntervalBound| into a \verb|JittableInputBound| at the input and output of any function to be transformed.
\subsubsection{Contributions} 
In this paper, we present a toolbox called \immrax\footnote{The most recent code for \immrax can be found at \url{https://github.com/gtfactslab/immrax}, and the documentation can be found at \url{https://immrax.readthedocs.io}.}, introducing several new function transformations to facilitate interval analysis and mixed monotone reachability analysis. 
These transforms are fully composable with existing JAX transformations, allowing the toolbox to support (i) Just-In-Time (JIT) Compilation for significant improvements in runtime versus the baseline, (ii) GPU parallelizability for rapid, accurate online reachable set estimation; (iii) Automatic Differentiation for learning relationships between reachable set outputs and input parameters. 
In Section~\ref{sec:inclusion}, we discuss the theory behind interval analysis and inclusion functions.
In particular, in Proposition~\ref{prop:MixedJacobian-Based} Part~\eqref{prop:MixedJacobian-Based:p1}, we provide a novel an analytical Jacobian-based bound for functions which is vital for capturing stabilizing interactions in closed-loop system analysis.
In Section~\ref{sec:embedding}, we discuss the theory behind using inclusion functions to build embedding systems which efficiently and scalably bound the output of a dynamical system under uncertainty.
In each Section, we present their corresponding implementations in \immrax as composable function transforms, keeping consistent with the rest of the JAX ecosystem.
Finally, in Section~\ref{sec:applications}, we demonstrate the usage of \immrax on several case studies including efficient reachable set estimation of a nonlinear system controlled by a neural network using GPU parallelization for partitioning, and finding locally optimal solutions to a robust closed-loop optimal control problem on a damped inverted pendulum using Automatic Differentiation.

\subsubsection{Notation} For $\ulx,\olx\in\R^n$, define the partial ordering $\ulx\leq\olx\iff\ulx_i\leq\olx_i$ for every $i=1,\dots,n$. Let $[\ulx,\olx] := \{x : \ulx\leq x \leq \olx\}$ denote a closed and bounded interval, and let $\IR^n$ be the set of all such intervals. The partial order $\leq$ on $\R^n$ induces the southeast order $\leq_{\text{SE}}$ on $\R^{2n}$, where $\smallconc{x}{\hat{x}}\leqse\smallconc{y}{\hat{y}}\iff x\leq y$ and $\hat{y}\leq\hat{x}$. Define the upper triangle $\Tgeq^{2n} := \{\smallconc{\ulx}{\olx}\in\R^{2n} : \ulx \leq \olx\}$, and note that $\IR^n\simeq\Tgeq^{2n}$. We denote this equivalence with $\left[\smallconc{\ulx}{\olx}\right] := [\ulx,\olx]$.
For $[\ula,\ola],[\ulb,\olb]\in\IR$ and $[\ulA,\olA]\in\IR^{m\times p},[\ulB,\olB]\in\IR^{p\times n}$,
\begin{enumerate}
    \item $[\ula,\ola] + [\ulb,\olb] := [\ula + \ulb, \ola + \olb]$ (also on $\IR^n$ element-wise);
    \item $[\ula,\ola]\cdot[\ulb,\olb] := [\min \{\ula\ulb,\ula\olb,\ola\ulb,\ola\olb\}, \max\{\ula\ulb,\ula\olb,\ola\ulb,\ola\olb\}] $;
    \item $([\ulA,\olA][\ulB,\olB])_{i,j} := \sum_{k=1}^{p} [\ulA_{i,k}, \olA_{i,k}]\cdot[\ulB_{k,j},\olB_{k,j}]$.
\end{enumerate}
For $x_1\in\R^{n_1}$, $x_2\in\R^{n_2}$, $\dots$, $x_m\in\R^{n_m}$, let $(x_1,x_2,\dots,x_m)\in\R^{n_1 + n_2 + \dots + n_m}$ denote their concatenation. 
For $f:\R^n\to\R^m$, let $df:\R^n\to\R^{m\times n}$ be its Jacobian, \emph{i.e.}, for $x'\in\R^n$, $df_{x'} = \frac{\partial f}{\partial x}\big|_{x=x'}$.

\section{Inclusion Module: Inclusion Function Transforms in JAX} \label{sec:inclusion}

The \verb|inclusion| module provides a streamlined interface to work with interval objects and inclusion functions.\footnote{At this time, \immrax does not bound floating point rounding errors.}

\subsection{Inclusion Functions}

Interval analysis provides a scalable, compositional approach to bound the output of a function along an interval input, and the key building block is the inclusion function.

\begin{defn}[Inclusion Function]
    Given a function $f:\R^n\to\R^m$, the function $\sfF = \smallconc{\ul{\sfF}}{\ol{\sfF}}:\Tgeq^{2n}\to\Tgeq^{2m}$ is an \emph{inclusion function} for $f$ if for every $x\in[\ulx,\olx]$,
    \[
        \ul{\sfF}(\ulx,\olx) \leq f(x) \leq \ol{\sfF}(\ulx,\olx).
    \]
    An inclusion function is \emph{monotone} if $[\ulx,\olx]\subseteq[\uly,\oly]$ implies 
    \[
        \ul{\sfF}(\uly,\oly) \leq \ul{\sfF}(\ulx,\olx) \leq f(x) \leq \ol{\sfF}(\ulx,\olx) \leq \ol{\sfF}(\uly,\oly).
    \]
    An inclusion function is \emph{thin} if for every $x\in\R^n$,
    \[
        \ul{\sfF}(x,x) = f(x) = \ol{\sfF}(x,x).
    \]
\end{defn}
\begin{rem}
    Notationally, we  use the upper triangular interpretation $\Tgeq^{2n}$ for convenience in Section~\ref{sec:embedding}. Most references instead think of inclusion functions as mappings on $\IR^n$. Given the equivalence between $\Tgeq^{2n}$ and $\IR^n$, we  use the notation $[\sfF]:\IR^n\to\IR^m$ to denote the equivalent interval input to output mapping.
\end{rem}

There are several methods for constructing inclusion functions. For some functions, it is possible to compute the minimal inclusion function, which returns the tightest possible output for a given interval input bound~\citep[Theorem 2.2]{AH-SJ-SC:23b}.

\begin{prop}[Minimal inclusion function] \label{prop:Minimal}
    Given a function $f:\R^n\to\R^m$, the unique, monotone and thin inclusion function returning the tightest bounds the image of $f$ on $[\ulx,\olx]$ is $\sfF = \smallconc{\ul\sfF}{\ol\sfF}$, where for every $i\in \{1,\ldots,n\}$,
    \[
    \ul{\sfF}_i(\ulx,\olx) = \inf_{x\in[\ulx,\olx]} f_i(x), \quad \ol{\sfF}_i(\ulx,\olx) = \sup_{x\in[\ulx,\olx]} f_i(x),
    \]
    Denote this as the \emph{minimal inclusion function} of $f$.
\end{prop}

Computing the minimal inclusion function is not generally viable. Instead, we provide several computationally efficient approaches to construct inclusion functions using known inclusion functions as building blocks. First, we present the natural inclusion function, which is the simplest technique~\citep{LJ-MK-OD-EW:01}, Proof in~\citep[Theorem 2.3]{AH-SJ-SC:23b}.

\begin{prop}[Natural inclusion function] \label{prop:Natural}
    Given a function $f:\R^n\to\R^m$, such that $f = f_1\circ f_2\circ\cdots \circ f_\ell$ is the composition of functions/operators $\{f_i\}_{i=1}^\ell$ with (monotone/thin) inclusion functions $\{\sfF_i\}_{i=1}^\ell$, the following is a (monotone/thin) inclusion function of $f$
    \[
    \sfF(\ulx,\olx) = (\sfF_1\circ \sfF_2\circ\cdots\circ \sfF_\ell)(\ulx,\olx).
    \]
    Denote this as the \emph{$\{\sfF_i\}_{i=1}^\ell$-natural inclusion function} of $f$.
\end{prop}

While the natural inclusion function provides a general approach, it is often overly conservative. Instead, if the function is differentiable, one can use a bound on the first order Taylor expansion of the function, which may provide better results in practice~\citep{LJ-MK-OD-EW:01}.

\begin{prop}[Jacobian-based inclusion function] \label{prop:Jacobian-based}
    $\\$ Consider a differentiable function $f:\R^n\to\R^m$, with an inclusion function $\sfJ$ for the Jacobian matrix $df$, \ie, $df_x\in [\sfJ(\ulx,\olx)]$ for every $x\in[\ulx,\olx]$. Then, any center $\overcirc{x}\in[\ulx,\olx]$ induces a valid inclusion function as follows
    \[
    [\sfF(\ulx,\olx)] = [\sfJ(\ulx,\olx)]([\ulx,\olx] - \overcirc{x}) + f(\overcirc{x}).
    \]
    Denote this as the $(\sfJ,\overcirc{x})$-\emph{Jacobian-based inclusion function} of $f$.
\end{prop}

By bounding each component of the vector input $x$ as separate variables, we can further reduce the overconservatism of the Jacobian-based approach. 
The following definition helps build the mixed Jacobian-based inclusion function.

\begin{defn}[Permutation] \label{def:Permutation}
    Given a dimension $n$, a $n$-\emph{permutation} $\sigma$ is a bijection of $\{1,\dots,n\}$ onto itself, characterized by a tuple of $n$ unique integers $1\leq\sigma(i)\leq n$. For an $n$-permutation $\sigma = (\sigma(1),\dots,\sigma(n))$, the $j$-th \emph{subpermutation} is $\sigma_j = (\sigma(1),\dots,\sigma(j))$. Define the \emph{replacement} $x_{\sigma_j:y}\in\R^n$ such that 
    \[(x_{\sigma_j:y})_i := \begin{cases}
        y_i & i\in\sigma_j \\
        x_i & i\notin\sigma_j
    \end{cases}.\]
\end{defn}

\begin{prop}[Mixed Jacobian-based inclusion function] \label{prop:MixedJacobian-Based}
    Consider a differentiable function $f:\R^n\to\R^m$, with an inclusion function $\sfJ$ for the Jacobian matrix $df$.
    Given a center $\overcirc{x}\in[\ulx,\olx]$ and an $n$-permutation $\sigma$, let $\sfM_\sigma^{\overcirc{x}}:\Tgeq^{2n}\to\Tgeq^{2(m\times n)}$ be defined such that for every $i=1,\dots,n$, the $\sigma(i)$-th column $[\sfM^{\overcirc{x}}_\sigma(\ulx,\olx)]_{\sigma(i)}:=[\sfJ(\overcirc{x}_{\sigma_i:\ulx},\overcirc{x}_{\sigma_i:\olx})]_{\sigma(i)}$.
    Then, the following statements hold:
    \begin{enumerate}[(i)]
        \item \label{prop:MixedJacobian-Based:p1} For every $x\in[\ulx,\olx]$,
        \[f(x)\in[\sfM_\sigma^{\overcirc{x}}(\ulx,\olx)](x - \overcirc{x}) + f(\overcirc{x});\]
        \item \label{prop:MixedJacobian-Based:p2} The function $\sfF:\Tgeq^{2n}\to\Tgeq^{2m}$, defined by
        \[[\sfF(\ulx,\olx)] = [\sfM^{\overcirc{x}}_\sigma(\ulx,\olx)]([\ulx,\olx] - \overcirc{x}) + f(\overcirc{x}),\]
        is an inclusion function for $f$, denoted as the $(\sfJ,\overcirc{x},\sigma)$-\emph{mixed Jacobian-based inclusion function} of $f$.
    \end{enumerate}
\end{prop}

\subsection{\texttt{immrax.inclusion} Module Implementation} 

In this subsection, we discuss the implementation of the interval analysis theory from the previous section in the submodule \verb|immrax.inclusion|. In particular, the various inclusion functions are implemented as function transforms composable with any existing JAX transformations.

\subsubsection{\texttt{Interval} Class}
In \immrax, intervals are implemented in the \verb|Interval| class, which is made up of two main attributes: \verb|lower| and \verb|upper|, which are \verb|jax.numpy.ndarray| objects of the same shape and dtype representing the lower and upper bound of the interval. \verb|Interval| is registered as a Pytree node, so JAX can internally handle any \verb|Interval| object as if it were a standard JAX type. This implementation differs from the \verb|IntervalBound| class from \verb|jax_verify|, which instead requires the user to manually swap between \verb|IntervalBound| objects and \verb|JittableInputBound| objects as needed.

\immrax provides several helper functions with input validation to safely construct and manipulate \verb|Interval| objects. For example, \verb|interval| creates an \verb|Interval| from a lower and upper bound; \verb|icentpert| creates an \verb|Interval| from a center $x_0$ and a vector $\epsilon$ from the center as $[x_0 - \epsilon, x_0 + \epsilon]$;  \verb|ut2i| converts a \verb|jax.Array| element of the upper triangle $\Tgeq^{2n}$ to its representation in $\IR^n$; and \verb|i2lu|, \verb|i2centpert|, \verb|i2ut| perform the inverse operations respectively.

\subsubsection{Inclusion Functions}

We provide minimal inclusion functions for basic \verb|jax.lax| primitives (and applicable class operations), such as \verb|add| (\verb|+|), \verb|sub| (\verb|-|), \verb|mul| (\verb|*|), \verb|div| (\verb|/|), \verb|pow| (\verb|**|), \verb|sin|, \verb|cos|, \verb|sqrt|, \verb|dot_general|. A full list of supported minimal inclusion functions can be found at the \immrax documentation, and this list will continue to grow.

The most versatile transform that \immrax provides is \verb|natif|, which implements the natural inclusion function from Proposition~\ref{prop:Natural}. Given a function \verb|f| acting on usual \verb|jax.Array| inputs, defined as a composition of primitives with defined inclusion functions, \verb|natif(f)| creates a new function replacing each primitive, or each functional building block, with their corresponding minimal inclusion function counterparts.
Internally, \immrax builds these inclusion functions by tracing the original \verb|f| into a \verb|ClosedJaxpr|, the JAX internal representation of the pure functional inputs, outputs, and intermediate operations. Then, the inclusion function \verb|F| is built by traversing the \verb|ClosedJaxpr|, replacing \verb|jax.Array| inputs with \verb|Interval|s and replacing primitives with their inclusion function.

In addition to \verb|natif|, \immrax provides transforms for the Jacobian-based inclusion function from Proposition~\ref{prop:Jacobian-based} as \verb|jacif| and the mixed Jacobian-based inclusion function as \verb|mjacif|. Internally, \immrax builds these inclusion functions by composing \verb|natif| with \verb|jacfwd|, which creates an inclusion function for the Jacobian matrix of \verb|f|. For example, a simple implementation of \verb|jacif| for single vector input functions of the form \verb|f(x)| is
\begin{verbatim}
def jacif (f) :
  df = immrax.natif(jax.jacfwd(f))
  def F (x:Interval, xc:jax.Array) -> Interval :
    return df(x) @ (x - xc) + f(xc)
  return F
\end{verbatim}
Note that the usage of \verb|@| here calls \verb|Interval.__matmul__|, which is explicitly defined in the \verb|inclusion| module as
\begin{verbatim}
Interval.__matmul__ = immrax.natif(jnp.matmul)
\end{verbatim}
In turn, \verb|jnp.matmul| uses the \verb|jax.lax.dot_general_p| primitive, for which the minimal inclusion function is provided.
In practice, the true implementations of \verb|jacif| and \verb|mjacif| work for functions of any number of inputs---and one can even specify multiple centers and/or multiple permutations using \verb|kwargs|, for which the minimum and maximum are taken accordingly.

Finally, all of these function transforms were carefully written to retain the ability to be composed with existing JAX transforms, such as \verb|jax.jit| for JIT compilation, \verb|jax.vmap| for parallelization, and \verb|jax.grad|/\verb|jax.jacfwd| for Automatic Differentiation. Note that the derivative of an interval or with respect to an interval is not directly a well defined object---instead, one can take the derivative of a real-valued function of an interval output, \emph{e.g.}, \verb|(out.upper - out.lower)|, the objective function for the pendulum in Section~\ref{sec:applications}, or simply the upper and/or lower bound.

The following example compares the inclusion functions generated from a call of \verb|natif|, \verb|jacif|, and \verb|mjacif|. 

\begin{exmp} \label{ex:incl_compared}
    For the function $f(x_1,x_2) = ((x_1 + x_2)^2, x_1 + x_2 + 2x_1x_2)$, we compare the natural, Jacobian-based, and mixed Jacobian-based inclusion functions on the input $[-0.1,0.1]\times[-0.1,0.1]$, for $\overcirc{x} = 0$ and $\sigma = (1,2)$, using \immrax:
\begin{verbatim}
f = lambda x : jnp.array([
  (x[0] + x[1])**2, x[0] + x[1] + 2*x[1]*x[2]])
Fnat = immrax.natif(f)
Fjac = immrax.jacif(f)
Fmix = immrax.mjacif(f)
x0 = immrax.icentpert(jnp.zeros(2), 0.1)
for F in [Fnat, Fjac, Fmix] :
  F(x0) # JIT Compile
  ret, times = utils.run_times(1000, F, x0)
\end{verbatim}
\begin{table}[h!]
    \centering
    \begin{tabular}{||c|c|c||}
        \hline
        \verb|F| & Output & Average Runtime \\
        \hline
        \verb|Fnat| & $[0,0.04]\times[-0.22,0.22]$ & $4.778\times10^{-5}$ \\
        \verb|Fjac| & $[-0.08,0.08]\times[-0.24,0.24]$ & $8.611\times10^{-5}$ \\
        \verb|Fmix| & $[-0.06,0.06]\times[-0.24,0.24]$ & $6.856\times10^{-5}$ \\
        \hline
    \end{tabular}
    \caption{Inclusion function outputs and runtimes for Example~\ref{ex:incl_compared}}
    \label{tab:my_label}
\end{table}
While the implementation of \verb|mjacif| might seem more complicated upon first glance, after being Just-In-Time Compiled, the performance is better than the standard \verb|jacif|. 
This is because a natural inclusion function computation takes on the order of at least twice as much as a standard computation
(upper and lower bound)---and while the Jacobian-based approach builds a full matrix of interval components using natural inclusion functions, the mixed Jacobian approach is able to reduce the number of interval computations by fixing some elements to the center $\overcirc{x}$, reducing to a standard singleton computation.
\end{exmp}

\section{Embedding Module: Mixed Monotone Embedding Systems in JAX} \label{sec:embedding}
In this section, we consider the theory and the implementation of continuous-time embedding systems in JAX, which provides an efficient and scalable method for bounding the reachable sets of dynamical systems.
\subsection{Mixed Monotone Embedding Systems}
Consider the nonlinear system
\begin{align} \label{eq:nlsys}
    \dot{x} = f(x,w),
\end{align}
where $x\in\R^n$ is the state of the system and $w\in\R^q$ is the disturbance input to the system. Assume that $\sfF$ is an inclusion function for $f$. Then, $\sfF$ induces the associated \emph{embedding system}
\begin{align} \label{eq:embsys}
\begin{aligned}
    \dot{\ulx}_i &= \ul{\sfE}_i(\ulx,\olx,\ulw,\olw) := \ul{\sfF}_i(\ulx,\olx_{i:\ulx},\ulw,\olw), \\
    \dot{\olx}_i &= \ol{\sfE}_i(\ulx,\olx,\ulw,\olw) := \ol{\sfF}_i(\ulx_{i:\olx},\olx,\ulw,\olw),
\end{aligned}
\end{align}
where $i\in \{1,\ldots,n\}$ and the new state $\smallconc{\ulx}{\olx}\in\Tgeq^{2n}$ evolves on the upper triangle, $\smallconc{\ulw}{\olw}\in\Tgeq^{2q}$. 
In the next proposition, we use a single trajectory of the embedding system to overapproximate the true reachable set of the system~\eqref{eq:nlsys}~\citep[Proposition 5]{SJ-AH-SC:23c}.
\begin{prop}[Reachability via embedding] \label{prop:reachset}
    Consider the system~\eqref{eq:nlsys} with an inclusion function $\sfF$ and its induced embedding system $\sfE$~\eqref{eq:embsys}. If $t\mapsto \smallconc{\ulx(t)}{\olx(t)}$ denotes the trajectory of $\sfE$ starting from initial condition $\smallconc{\ulx_0}{\olx_0}\in\Tgeq^{2n}$ at $t_0$ with disturbance $\smallconc{\ulw}{\olw}\in\Tgeq^{2n}$, then for every $t\geq t_0$, 
    $x(t) \in [\ulx(t),\olx(t)],$
    where $t\mapsto x(t)$ is the trajectory of~\eqref{eq:nlsys} from initial condition $x_0\in[\ulx_0,\olx_0]$.
\end{prop}
The problem of evaluating an infinite number of trajectories for the reachable set is replaced with overapproximated interval bounds using a single trajectory of the embedding system, which provides an efficient, scalable approach for online reachable set estimation.
\begin{rem}
    If $\sfF$ is a monotone inclusion function, then the induced embedding system $\sfE$ is a monotone dynamical system~\citep{DA-EDS:03} with respect to the southeast partial order $\leqse$.
    If $\sfF$ is additionally a thin inclusion function, then the approach is equivalent to the decomposition-based approach from~\cite{SC-MA:15b}. Thinness ensures that the decomposition function 
    \[d_i(x,\hat{x},w,\hat{w}) := \begin{cases}
        \ul{\sfF}_i(x,\hat{x}_{i:x},w,\hat{w}) & x \leq \hat{x},\,w\leq \hat{w} \\
        \ol{\sfF}_i(\hat{x}_{i:x},x,\hat{w},w) & \hat{x} \leq x,\,\hat{w}\leq w \\
    \end{cases}\]
    satisfies $d(x,x,w,w) = f(x,w)$. 
\end{rem}

\subsection{\texttt{immrax.embedding} Module Implementation}

Similar to the usage of the inclusion function transforms, \immrax provides transforms on dynamical systems, generating dynamical embedding systems. First, one defines a \verb|System| object, which evolves on a state $\verb|x|\in\R^n$ and defines the vector field $\verb|f|:\R^n\times\cdots \to \R^n$. Here, $\cdots$ represents any number of additional inputs to the system.
Given a \verb|System| object \verb|sys|, and an inclusion function \verb|F| for the dynamics \verb|sys.f|, the transform \verb|ifemb(sys, F)| returns an \verb|EmbeddingSystem| object whose dynamics are constructed using~\eqref{eq:embsys} on the inclusion function \verb|F|.
For convenience, the transforms \verb|natemb(sys)|, \verb|jacemb(sys)|, and \verb|mjacemb(sys)| automatically construct the \verb|EmbeddingSystem| induced by the natural inclusion function (\verb|natif(sys.f)|), the Jacobian-based inclusion function (\verb|jacif(sys.f)|), and the Mixed Jacobian-based inclusion function (\verb|mjacif(sys.f)|).

\section{Applications}\label{sec:applications}

We demonstrate the usage of \immrax through reachability on a nonlinear vehicle controlled by a neural network and robust closed-loop control synthesis on a pendulum.\footnote{All experiments were performed on a computer running Kubuntu 22.01, with a Ryzen 5 5600X, Nvidia RTX 3070, and 32 GB of RAM.}

\subsection{GPU Acceleration for Neural Network Feedback Loops}

In this example, we reimplement the interaction-aware first-order inclusion function from our previous work \texttt{ReachMM}~\citep{SJ-AH-SC:23c} using \immrax. 
We compare to the nonlinear bicycle model~\citep[Section VII.A]{SJ-AH-SC:23c}---and the full implementation details can be found in the correponding Jupyter notebook in the \immrax documentation.
In Table~\ref{tab:ARCH_table}, we compare the runtimes (after JIT compilation) across several different numbers of initial partitions on the CPU/GPU, as well as to a similar (hybrid mode) implementation in \verb|ReachMM|~\citep{SJ-AH-SC:23c}.
The compiled \immrax implementation sees substantial improvement in the runtime on the CPU, allowing it to perform the higher fidelity Tsit5 algorithm in a similar runtime as Euler integration on \verb|ReachMM|. 
Additionally, while the GPU performance on a single initial partition is worse than the CPU, as the number of initial partitions increase, the benefits of the parallelization is clear as it can compute, \emph{e.g.}, $625$ partitions in less than $10\times$ the runtime of a single partition on the CPU.

\begin{table}[]
    \centering \label{tab:vehicle}
    \begin{tabular}{||c | c c c c c ||}
        \hline
        \multirow{2}{*}{\# Part.} & \texttt{ReachMM} & \multicolumn{2}{c}{\immrax (Euler)} & \multicolumn{2}{c ||}{\immrax (tsit5)} \\
        & (Euler) & CPU & GPU & CPU & GPU \\
        \hline\hline
        $1^4 = 1$    & $.0476$ & $.0112$ & $.0178$ & $.0649$ & $.0983$ \\
        $2^4 = 16$   & $.690$  & $.143$  & $.0207$ & $.856$  & $.112$ \\
        $3^4 = 81$   & $3.44$  & $.627$  & $.0306$ & $3.86$  & $.187$ \\
        $4^4 = 256$  & $11.0$  & $1.44$  & $.0489$ & $8.87$  & $.302$ \\
        $5^4 = 625$  & $27.1$  & $4.60$  & $.095$  & $27.9$  & $.588$ \\
        $6^4 = 1296$ & $55.8$  & $11.1$  & $.198$  & $67.1$  & $1.13$ \\
        \hline
    \end{tabular}
    \caption{Summary of average runtimes (over 10 runs) in seconds for the bicycle model on different numbers of initial partitions.}
    \label{tab:ARCH_table}
\end{table}

\subsection{Automatic Differentiation for Robust Optimal Control}

In this example, we use Automatic Differentiation to solve a robust optimal control problem in the embedding space of a nonlinear pendulum.
In particular, we provide and discuss the relevant \immrax code needed to build an objective function with robust constraints, automatically create and compile functions evaluating their gradients, Jacobians, and Hessians, and finally setup an IPOPT minimization problem to find a locally optimal solution.

Consider the dynamics of a forced, damped pendulum 
\begin{align}
    ml^2\ddot{\theta} + b\dot{\theta} + mgl\sin\theta = \tau,
\end{align}
with $m=0.15\mathrm{kg}$, $l=0.5\mathrm{m}$, $b=0.1\mathrm{N\cdot m \cdot s}$, and $g=9.81\mathrm{m}/\mathrm{s}^2$.
The torque $\tau := (1 + w)u$, where $u\in\R$ is the desired torque input and $w\in[\ulw,\olw] := [-0.02,0.02]$ is a bounded multiplicative disturbance on the control input. We implement this as a $2$-state system with $x := (\theta, \dot{\theta})$,
\begin{align}
    \dot{x} = 
     f(x,u,w) =  
    \conc{x_2}{\frac{(1 + w)u - bx_2}{ml^2} - \frac{g}{l}\sin x_1}
\end{align}
This is implemented in \immrax as a \texttt{System}, with the specified dynamics written using \texttt{jax.numpy}.
\begin{verbatim}
import jax.numpy as jnp
import immrax
class Pendulum (immrax.System) :
  def __init__(self, m=0.15, l=0.5, b=0.1) :
    self.evolution = 'continuous'
    self.xlen = 2
    self.m = m; self.l = l; self.b = b
  def f (self, t, x, u, w) :
    return jnp.array([ x[1],
      (((1 + w[0])*u[0] - self.b*x[1]) / 
        (self.m * self.l**2)) 
        - (g/self.l)*jnp.sin(x[0]) ])
sys = Pendulum()
\end{verbatim}
In the preceding code, we specified two key properties for the \texttt{Pendulum}: \verb|evolution| tells \verb|immrax| that the system is in continuous time, which is needed to build the proper continuous embedding system, and \verb|xlen| tells \verb|immrax| the length of the state vector.

We seek to find a finite-time closed-loop optimal control policy $\pi:[0,T]\times\R^n\to\R$ to swing up the pendulum to an \emph{a priori} safe region at the top.
We consider linear feedback control policies of the form $\pi(t,x) := K(x(t) - x_{\mathrm{nom}}(t)) + u_{\mathrm{ff}}(t)$, where $K$ is a time invariant linear closed-loop stabilizing term to help counter the disturbance, $u_{\mathrm{ff}}:[0,T]\to\R$ is a feedforward control policy, and $x_\mathrm{nom}:[0,T]\to\R^n$ is the nominal trajectory of the deterministic system under the feedforward control law $u_\mathrm{ff}$ with known disturbance mapping $w_{\mathrm{nom}}:[0,T]\to\R$.
The closed-loop system is thus
\begin{align} \label{eq:pendulumcldyn}
    \dot{x} = f(x,\pi(t,x),w) = f^\pi (t, x, w).
\end{align}
Consider the following function
\begin{align}
    [\sfF^\pi(t,\ulx,\olx,\ulw,\olw)] := &\, ([\sfM_x] + [\sfM_u]K)([\ulx,\olx] - x_\mathrm{nom}(t)) \nonumber \\
    & + [\sfM_w]([\ulw,\olw] - w_\mathrm{nom}(t))  \label{eq:pendulumif} \\
    & + f(x_\mathrm{nom}(t),u_{\mathrm{ff}}(t),w_\mathrm{nom}(t)), \nonumber
\end{align}
with $[\sfM_x\ \sfM_u\ \sfM_w] := [\sfM_\sigma^{\xi_\mathrm{nom}(t)}(\ulx,\olx,u_\mathrm{ff}(t),u_\mathrm{ff}(t),\ulw,\olw)]$, where $\sfM$ is defined as Proposition~\ref{prop:MixedJacobian-Based} for the map $\hat{f} : \R^{n+p+q} \to \R^n$ such that $\hat{f}((x,u,w)) := f(x,u,w)$, for some $(n+p+q)$-permutation $\sigma$, and $\xi_\mathrm{nom}(t) := (x_\mathrm{nom}(t),u_\mathrm{ff}(t),w_\mathrm{nom}(t))$.
This is a valid inclusion function for the closed-loop dynamics $f^\pi$~\eqref{eq:pendulumcldyn} (proof uses Proposition~\ref{prop:MixedJacobian-Based} Part~\eqref{prop:MixedJacobian-Based:p1}).
We use the \verb|mjacM| transform to implement the inclusion function in \immrax as follows 
\begin{verbatim}
sys_mjacM = immrax.mjacM(sys.f)
def F (t, x, w, K, nominal) :
  tc, xc, uc, wc = nominal
  iuc = immrax.interval(uc)
  iK = immrax.interval(K)
  Mt, Mx, Mu, Mw = sys_mjacM(t, x, 
    iuc, w, centers=(nominal,))[0]
  return ((Mx + Mu @ iK) @ (x - xc)
    + Mw @ (w - wc) + sys.f(tc, xc, uc, wc))
embsys = immrax.ifemb(sys, F)
\end{verbatim}
The final step creates the embedding system \verb|embsys|, \emph{i.e.} $\sfE$ from~\eqref{eq:embsys} induced by the inclusion function $\sfF^\pi$~\eqref{eq:pendulumif}.
Using the embedding system, we would like to solve the following robust optimal control problem,
\begin{align} \label{eq:pendopt}
    \min_{u_\mathrm{ff}, K} &\sum_{i=1}^N |u_{\mathrm{ff}}(t_i)|^2 + \|K\|_F^2 + \sum_{i=1}^N \|\olx(t_i) - \ulx(t_i)\|_2^2 \nonumber \\
    \text{s.t.}& \ \  \ulx_f \leq \ulx(t_j),\ \ \olx(t_j) \leq \olx_f, \ \ j=N_e,\dots,N, \ \\ 
    & \ulx(0) = \olx(0) = (0, 0), \nonumber \\
    & \smallconc{\ulx(t_{i+1})}{\olx(t_{i+1})} = \smallconc{\ulx(t_i)}{\olx(t_i)} + \Delta t \sfE(t_i,\ulx(t_i),\olx(t_i),\ulw,\olw), \nonumber 
\end{align}
where the embedding dynamics $\sfE$ are discretized using Euler integration with step size $\Delta t$.
The first and second terms of the objective are typical quadratic conditioning of the decision variables. The third term is a regularization factor intended to help curb the expansion of the gap between the upper and lower bound, which empirically helps the optimization problem converge to a feasible solution.
Finally, in the inequality constraints, we require that the pendulum reach a target set $[\ulx_f,\olx_f]$ and stay within these constraints for $t\in[t_{N_e},t_N]$.

To setup the minimization problem in IPOPT, we first implement a function called \verb|rollout_cl_embsys|, which uses \texttt{jax.lax.scan} to perform Euler integration on the dynamics for a given control sequence $(u_\mathrm{ff}(t_i))_{i=1}^N$ and gain matrix $K$, returning the state sequence $(x_{t_i})_{i=1}^N$.
\begin{verbatim}
def rollout_cl_embsys (u) :
  u, K = split_u(u)
  def f_euler (xt, ut) :
    xtut, xnomt = xt
    xtutp1 = xtut + dt*embsys.E(0., xtut, w, K, 
        (jnp.array([0.]), xnomt, 
            jnp.array([ut]), jnp.array([0.])))
    xnomtp1 = xnomt + dt*sys.f(0., xnomt, 
        jnp.array([ut]), jnp.array([0.]))
    return ((xtutp1,xnomtp1), xtutp1)
  _,x = jax.lax.scan(f_euler, (x0ut,x0cent), u)
  return x
\end{verbatim}
Note that the input $\texttt{u}\in\R^{N+2}$ holds all decision variables, \emph{i.e.}, $\texttt{u} := (u_{t_1}, u_{t_2},\dots, u_{t_N}, K_{1,1}, K_{1,2})$.
To compute the nominal trajectory $x_\mathrm{nom}$, we simulate the undisturbed system ($w_\mathrm{nom} = 0$).
Next, we implement the objective,
\begin{verbatim}
def obj (u) :
  x = rollout_cl_embsys(u)
  return (jnp.sum(u**2) 
    + jnp.sum((x[:,2:] - x[:,:2])**2))
\end{verbatim}
The final set is implemented as $\verb|con_ineq(u)| \geq 0$,
\begin{verbatim}
xf = immrax.icentpert([jnp.pi,0.],
                      [10*(jnp.pi/360),.1])
xfl, xfu = immrax.i2lu(xf)
def con_ineq (u) :
  x = rollout_cl_embsys(u)
  return jnp.concatenate((
    (x[Ne:,:2] - xfl).reshape(-1), 
    (xfu - x[Ne:,2:]).reshape(-1)))
\end{verbatim}

\begin{figure}
    \centering
    \includegraphics[width=0.32\columnwidth]{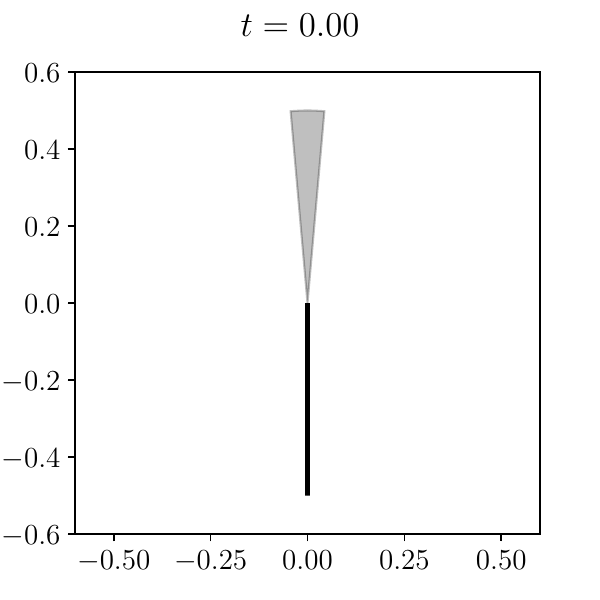}
    \includegraphics[width=0.32\columnwidth]{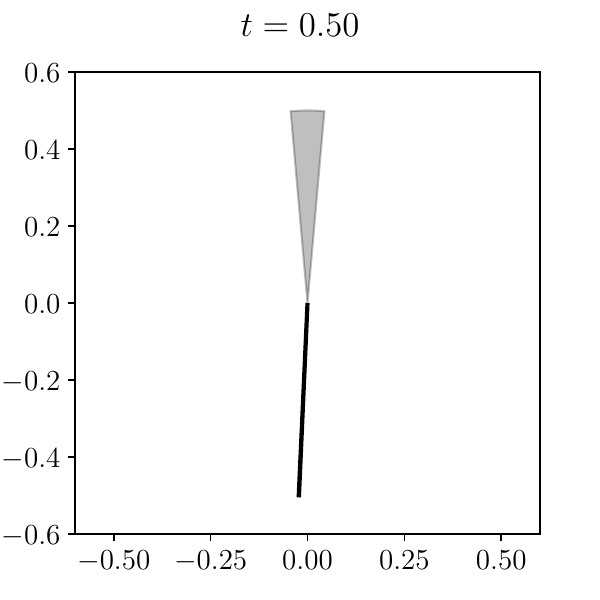}
    \includegraphics[width=0.32\columnwidth]{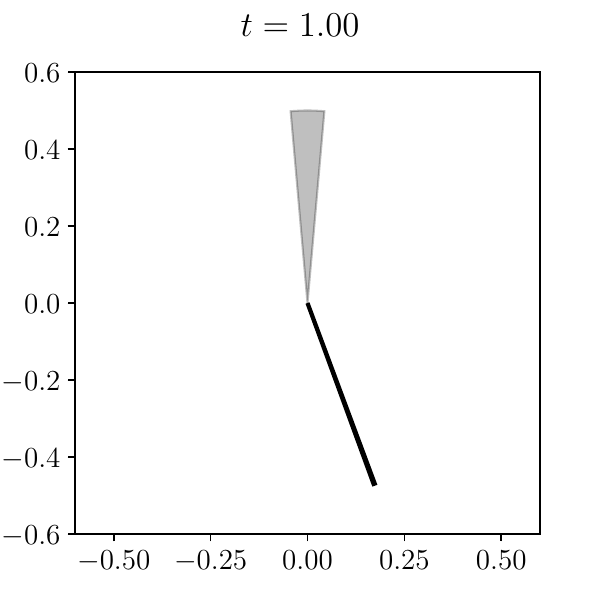}
    \includegraphics[width=0.32\columnwidth]{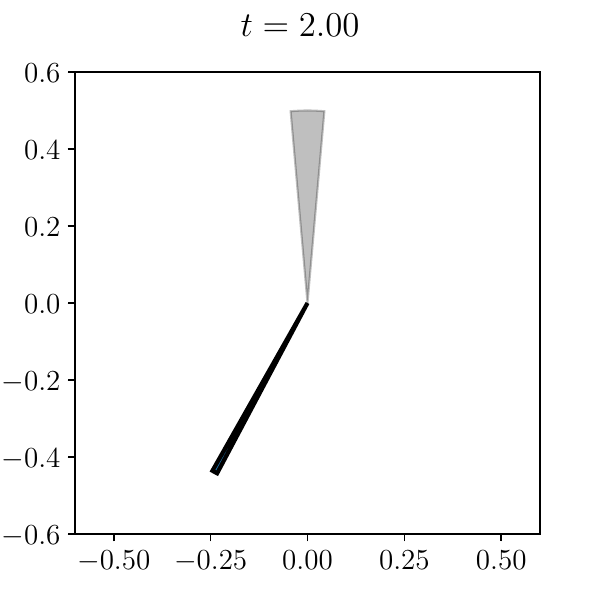}
    \includegraphics[width=0.32\columnwidth]{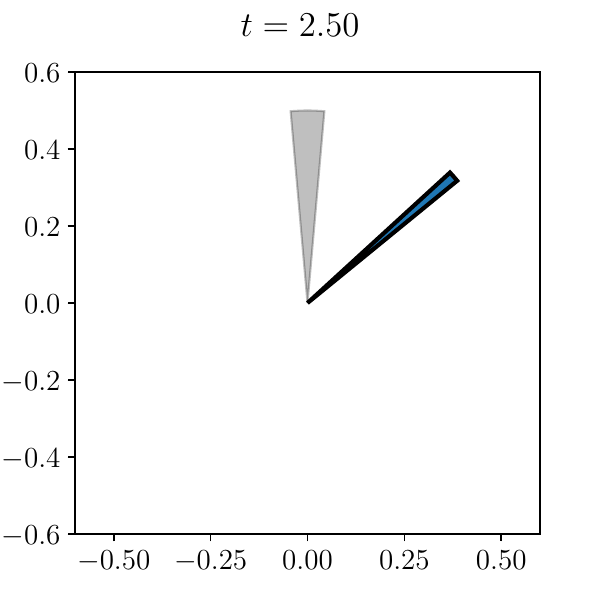}
    \includegraphics[width=0.32\columnwidth]{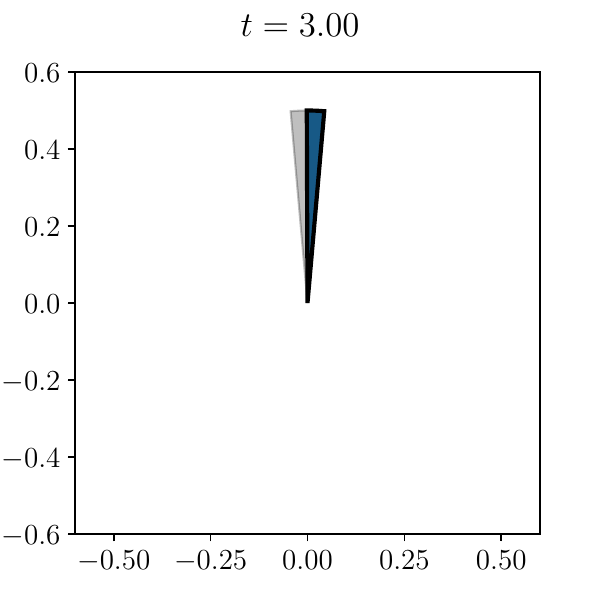}
    \caption{The swinging trajectory of the pendulum embedding system induced by~\eqref{eq:pendulumif} controlled by the closed-loop control policy generated by the problem~\eqref{eq:pendopt} is visualized for various time instances. The angle $[\ul{\theta},\ol{\theta}]$ is represented as a wedge, where blue represents the interval of possible angles. The gray wedge represents the desired final state of the pendulum.}
    \label{fig:pend_swinging}
\end{figure}

Next, we use JAX's autodiff transforms to automatically create functions to compute the objective's gradient and Hessian, as well as the Jacobian and Hessian vector product of the constraints with respect to Lagrange multipliers.
\begin{verbatim}
obj_grad = jax.grad(obj)
obj_hess = jax.jacfwd(jacrev(obj))
con_ineq_jac = jax.jacfwd(con_ineq)
def con_ineq_hessvp (u, v) :
  def hessvp (u) :
    _, hvp = jax.vjp(con_ineq, u)
    return hvp(v)[0]
  return jax.jacrev(hessvp)(u)
\end{verbatim}
After JIT compiling all of these functions, we use \verb|cyipopt| with the MA57 linear solver to find a feasible solution.
\begin{verbatim}
cons = [{'type': 'ineq', 'fun': con_ineq, 
  'jac': con_ineq_jac, 'hess': con_ineq_hessvp}]
res = minimize_ipopt(obj, jac=obj_grad, 
  hess=obj_hess, x0=u0, constraints=cons, 
  options=ipopt_opts)
\end{verbatim}

The setup and compilation steps took $147.91$ seconds, and IPOPT was run for $100$ iterations, taking $2.60$ seconds and satisfying all constraints with a tolerance of $-1.40\times10^{-4}$.
The resulting trajectory is visualized in Figure \ref{fig:pend_swinging}. 

\section{Conclusions}
In this paper, we presented a differentiable and parallelizable framework for interval analysis in JAX.
We applied this framework to two case studies demonstrating the toolbox's potential for efficient closed-loop reachability analysis and robust controller design.
Future work will involve certified robust training of neural network controllers.

\bibliography{SJ,SJ2}             %

\begin{thebibliography}{11}
\providecommand{\natexlab}[1]{#1}
\providecommand{\url}[1]{\texttt{#1}}
\providecommand{\urlprefix}{URL }
\expandafter\ifx\csname urlstyle\endcsname\relax
  \providecommand{\doi}[1]{doi:\discretionary{}{}{}#1}\else
  \providecommand{\doi}{doi:\discretionary{}{}{}\begingroup
  \urlstyle{rm}\Url}\fi

\bibitem[{Althoff(2015)}]{CORA}
Althoff, M. (2015).
\newblock An introduction to {CORA} 2015.
\newblock In \emph{Proc. of the 1st and 2nd Workshop on Applied Verification
  for Continuous and Hybrid Systems}, 120--151. EasyChair.
\newblock \doi{10.29007/zbkv}.

\bibitem[{Angeli and Sontag(2003)}]{DA-EDS:03}
Angeli, D. and Sontag, E.D. (2003).
\newblock Monotone control systems.
\newblock \emph{IEEE Transactions on Automatic Control}, 48(10), 1684--1698.
\newblock \doi{10.1109/TAC.2003.817920}.

\bibitem[{Bogomolov et~al.(2019)Bogomolov, Forets, Frehse, Potomkin, and
  Schilling}]{SB-etal:19}
Bogomolov, S., Forets, M., Frehse, G., Potomkin, K., and Schilling, C. (2019).
\newblock {JuliaReach}: a toolbox for set-based reachability.
\newblock In \emph{Proc. of the 22nd International Conference on Hybrid
  Systems: Computation and Control}, 39--44.

\bibitem[{Bradbury et~al.(2018)Bradbury, Frostig, Hawkins, Johnson, Leary,
  Maclaurin, Necula, Paszke, Vander{P}las, Wanderman-{M}ilne, and
  Zhang}]{jax2018github}
Bradbury, J., Frostig, R., Hawkins, P., Johnson, M.J., Leary, C., Maclaurin,
  D., Necula, G., Paszke, A., Vander{P}las, J., Wanderman-{M}ilne, S., and
  Zhang, Q. (2018).
\newblock {JAX}: composable transformations of {P}ython+{N}um{P}y programs.
\newblock \urlprefix\url{http://github.com/google/jax}.

\bibitem[{Coogan and Arcak(2015)}]{SC-MA:15b}
Coogan, S. and Arcak, M. (2015).
\newblock Efficient finite abstraction of mixed monotone systems.
\newblock In \emph{Proceedings of the 18th International Conference on Hybrid
  Systems: Computation and Control}, 58--67.

\bibitem[{Fan et~al.(2020)Fan, Huang, Chen, Li, and Zhu}]{JF-CH-XC-WL-ZQ:20}
Fan, J., Huang, C., Chen, X., Li, W., and Zhu, Q. (2020).
\newblock {ReachNN*}: A tool for reachability analysis of neural-network
  controlled systems.
\newblock In \emph{International Symposium on Automated Technology for
  Verification and Analysis}, 537--542. Springer.

\bibitem[{Harapanahalli et~al.(2023)Harapanahalli, Jafarpour, and
  Coogan}]{AH-SJ-SC:23b}
Harapanahalli, A., Jafarpour, S., and Coogan, S. (2023).
\newblock A toolbox for fast interval arithmetic in \texttt{numpy} with an
  application to formal verification of neural network controlled system.
\newblock In \emph{2nd ICML Workshop on Formal Verification of Machine
  Learning}.

\bibitem[{Jafarpour et~al.(2023)Jafarpour, Harapanahalli, and
  Coogan}]{SJ-AH-SC:23c}
Jafarpour, S., Harapanahalli, A., and Coogan, S. (2023).
\newblock Efficient interaction-aware interval analysis of neural network
  feedback loops.
\newblock \emph{arXiv preprint arXiv:2307.14938}.

\bibitem[{Jaulin et~al.(2001)Jaulin, Kieffer, Didrit, and
  Walter}]{LJ-MK-OD-EW:01}
Jaulin, L., Kieffer, M., Didrit, O., and Walter, {\'E}. (2001).
\newblock \emph{Applied Interval Analysis}.
\newblock Springer London.

\bibitem[{Shen and Scott(2017)}]{KS-JKS:17}
Shen, K. and Scott, J.K. (2017).
\newblock Rapid and accurate reachability analysis for nonlinear dynamic
  systems by exploiting model redundancy.
\newblock \emph{Computers \& Chemical Engineering}, 106, 596--608.
\newblock ESCAPE-26.

\bibitem[{Wang et~al.(2023)Wang, Zhou, Fan, Wang, Li, Chen, Huang, Li, and
  Zhu}]{YW-etal:23}
Wang, Y., Zhou, W., Fan, J., Wang, Z., Li, J., Chen, X., Huang, C., Li, W., and
  Zhu, Q. (2023).
\newblock {Polar-Express}: Efficient and precise formal reachability analysis
  of neural-network controlled systems.
\newblock \emph{IEEE Transactions on Computer-Aided Design of Integrated
  Circuits and Systems}.

\end{thebibliography}

\appendix

\section{Additional Figures} \label{apx:sec:additionalfigures}

\begin{figure}[h] \label{fig:vehicle}
    \centering
    \includegraphics[width=\columnwidth, trim={0cm 0cm 0 0}, clip]{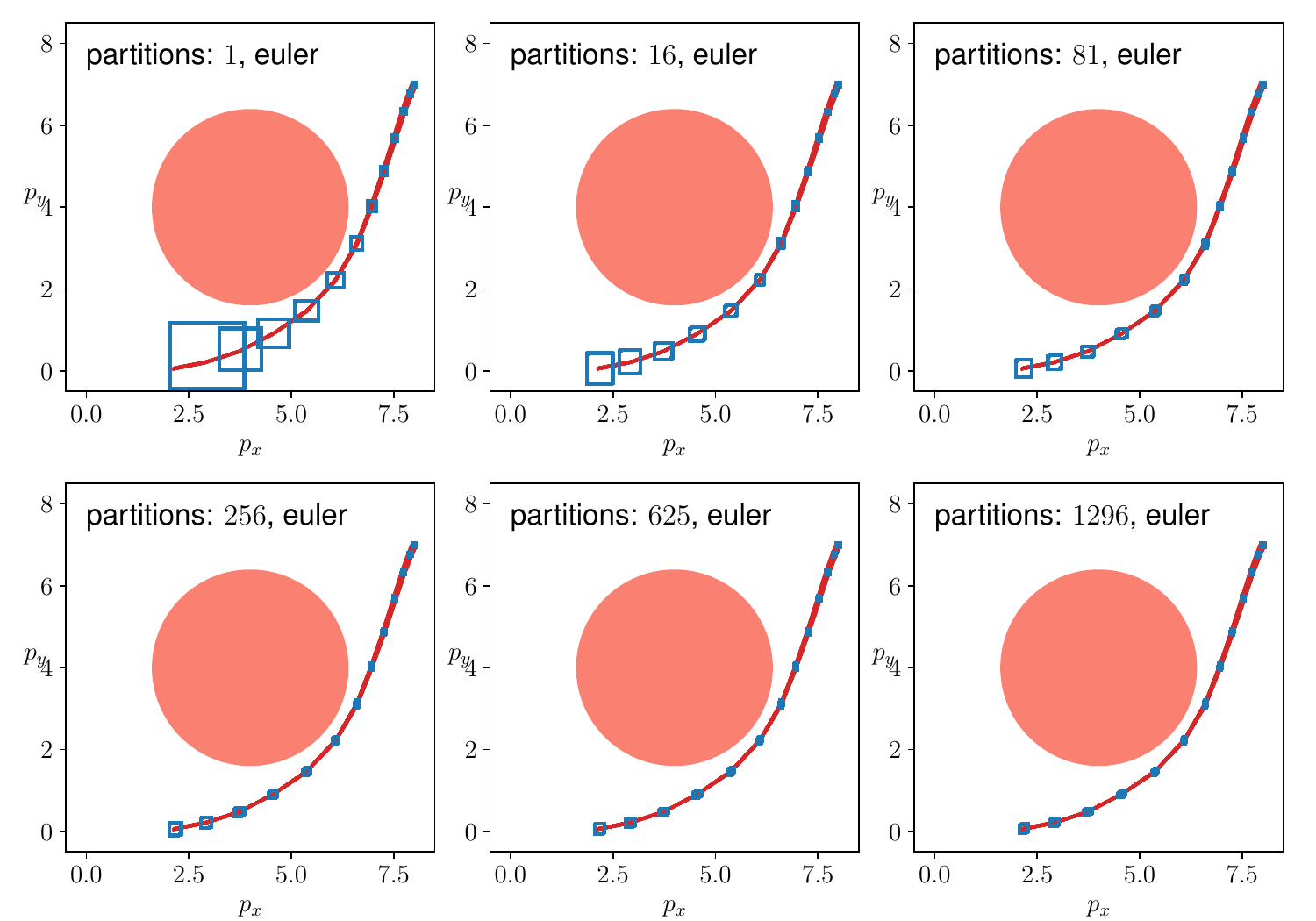}
    \caption{The reachable set over-approximations computed by simulating the embedding system from~\citep{SJ-AH-SC:23c} using Euler integration are visualized in light blue. The initial set $[7.95,8.05]\times[6.95,7.05]\times[-\frac{2\pi}{3} - 0.01, -\frac{2\pi}{3} + 0.01]\times[1.99,2.01]$ is divided into different numbers of partitions. $100$ Monte Carlo trajectories are pictured in dark red. In all cases, the vehicle is certified to avoid the obstacle pictured in light red, with varying degrees of accuracy to the true reachable set.}
    \label{fig:pend_tplots}
\end{figure}

\begin{figure}[h]
    \centering
    \includegraphics[width=0.9\columnwidth]{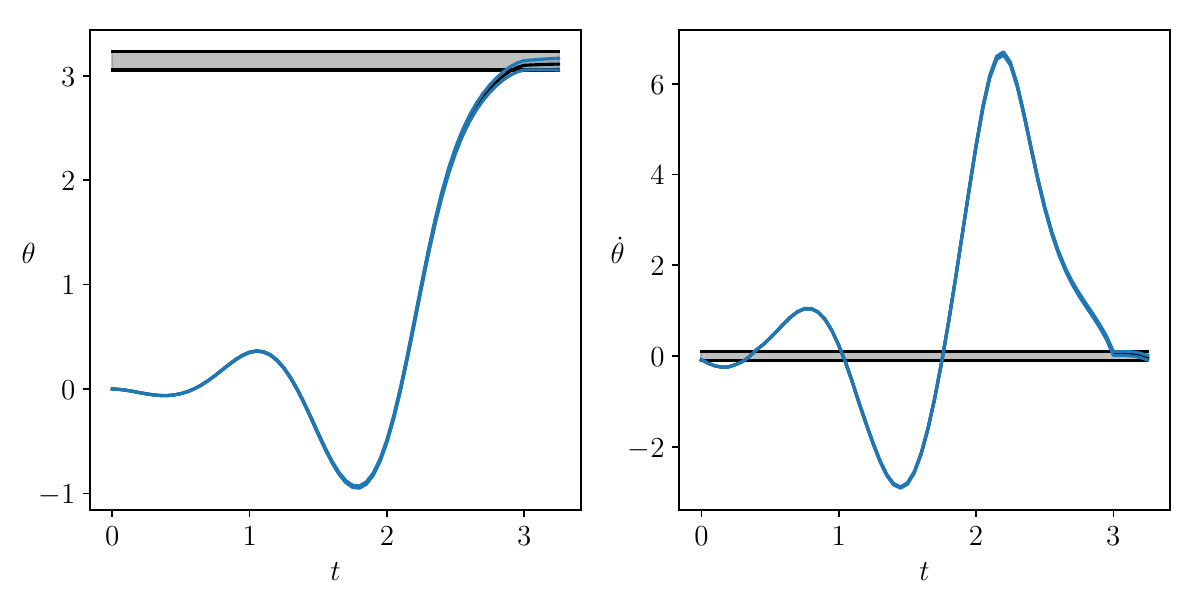}
    \caption{
    The swinging trajectory of the embedding system induced by~\eqref{eq:pendulumif} controlled by the closed-loop control policy generated by the optimization problem~\eqref{eq:pendopt} 
    is plotted versus time. \textbf{Left:} The angle $[\ul{\theta},\ol{\theta}]$ vs. $t$ in seconds (blue), with the terminal set constraint $\theta\in[\pi-\frac{10\pi}{360},\pi+\frac{10\pi}{360}]$ (gray). 
    \textbf{Right:} The angluar velocity $[\ul{\dot{\theta}},\ol{\dot{\theta}}]$ vs. $t$ in seconds, with the terminal set constraint $\dot{\theta}\in[-0.1,0.1]$ (gray).
    }
    \label{fig:pend_tplots}
\end{figure}

\section{Proof of Closed-Loop Pendulum Inclusion Function} \label{apx:sec:proofclpend}
In this appendix, we prove that~\eqref{eq:pendulumif} is an inclusion function for the system~\eqref{eq:pendulumcldyn}.
Consider the nonlinear system 
\begin{align} \label{eq:pendapx:nlsys}
    \dot{x} = f(x,u,w),
\end{align}
where $f:\R^n\times\R^p\times\R^q\to\R^n$ is a parameterized vector field.
Consider a piecewise continuous feedforward control curve $[0,T]\ni t \mapsto u_\mathrm{ff}(t)\in\R^p$,
a piecewise continuous disturbance trajectory $[0,T]\ni t\mapsto w_\mathrm{nom}(t)\in[\ulw,\olw]$ and some initial condition $x_0\in\R^n$. Let $[0,T]\ni t\mapsto x_\mathrm{nom}(t)\in\R^n$ denote the corresponding trajectory of~\eqref{eq:pendapx:nlsys} from initial condition $x_0$, under control mapping $u_\mathrm{ff}$ and disturbance mapping $w_\mathrm{nom}$. For some gain matrix $K\in\R^{p\times n}$, define the feedback control policy $\pi:[0,T]\times\R^n\to\R^p$, such that $\pi(t,x) = K(x - x_\mathrm{nom}(t)) + u_\mathrm{ff}(t)$. Denote the closed-loop system as
\begin{align} \label{eq:pendapx:clsys}
    \dot{x} = f(x,\pi(t,x),w) =: f^\pi(t,x,w).
\end{align}

Fix some $t\in[0,T]$, $[\ulx,\olx]\in\IR^n$, and $[\ulw,\olw]\in\IR^q$.
One can consider the vector field $f:\R^n\times\R^p\times\R^q\to\R^n$ as a mapping $\hat{f}:\R^{n+p+q}\to\R^n$. 
Let $(x,u,w):=\xi$, $(\overcirc{x},\overcirc{u},\overcirc{w}) := \overcirc{\xi}$, and define $\ul{\xi} := (\ulx,u_\mathrm{ff}(t),\ulw)$ and $\ol{\xi} := (\olx,u_\mathrm{ff}(t),\olw)$.
Then, for any center $\overcirc{\xi}\in[\ul{\xi},\ol{\xi}]$ and any permutation $\sigma$ on $(n+p+q)$, Proposition~\ref{prop:MixedJacobian-Based}\eqref{prop:MixedJacobian-Based:p1} implies that for every $\xi\in[\ul{\xi},\ol{\xi}]$,
\begin{align} 
    \hat{f}(\xi) \in [\sfM_\sigma^{\overcirc{\xi}}(\ul{\xi},\ol{\xi})] (\xi - \overcirc{\xi}) + \hat{f}(\overcirc{\xi}).
\end{align}
With the definition $[\sfM_x^{\overcirc{\xi}} \ \sfM_u^{\overcirc{\xi}} \ \sfM_w^{\overcirc{\xi}}] := [\sfM_\sigma^{\overcirc{\xi}}]$, the previous statement is equivalent to
\begin{align}
\begin{aligned}
    f(x,u,&\,w) \in [\sfM_x^{\overcirc{\xi}}(\ulx,\olx,u_\mathrm{ff}(t),u_\mathrm{ff}(t),\ulw,\olw)] (x - \overcirc{x}) \\
    &+ [\sfM_u^{\overcirc{\xi}}(\ulx,\olx,u_\mathrm{ff}(t),u_\mathrm{ff}(t),\ulw,\olw)] (u - u_\mathrm{ff}(t)) \\
    &+ [\sfM_w^{\overcirc{\xi}}(\ulx,\olx,u_\mathrm{ff}(t),u_\mathrm{ff}(t),\ulw,\olw)] (w - \overcirc{w}) \\
    &+ f(\overcirc{x},u_\mathrm{ff}(t),\overcirc{w}).
\end{aligned}
\end{align}
Let $\overcirc{x} := x_\mathrm{nom}(t)$, $\overcirc{w} := w_\mathrm{nom}(t)$, $u := \pi(t,x) = K(x - x_\mathrm{nom}(t)) + u_\mathrm{ff}(t)$, and $\xi_\mathrm{nom}(t) := (x_\mathrm{nom}(t), u_\mathrm{ff}(t), w_\mathrm{nom}(t))$. 
If $x_\mathrm{nom}(t)\in[\ulx,\olx]$ and $w_\mathrm{nom}(t)\in[\ulw,\olw]$,
\begin{align}
\begin{aligned}
    f^\pi&(t,x,w) \in [\sfM_x^{\xi_\mathrm{nom}(t)}(\ulx,\olx,u_\mathrm{ff}(t),u_\mathrm{ff}(t),\ulw,\olw)] (x - x_\mathrm{nom}(t)) \\
    &+ [\sfM_u^{\xi_\mathrm{nom}(t)}(\ulx,\olx,u_\mathrm{ff}(t),u_\mathrm{ff}(t),\ulw,\olw)] (K(x - x_\mathrm{nom}(t))) \\
    &+ [\sfM_w^{\xi_\mathrm{nom}(t)}(\ulx,\olx,u_\mathrm{ff}(t),u_\mathrm{ff}(t),\ulw,\olw)] (w - w_\mathrm{nom}(t)) \\
    &+ f(x_\mathrm{nom}(t),u_\mathrm{ff}(t),w_\mathrm{nom}(t)).
\end{aligned}
\end{align}
Combining terms,
\begin{align}
\begin{aligned}
    f^\pi&(t,x,w) \in  ([\sfM_x^{\xi_\mathrm{nom}(t)}(\ulx,\olx,u_\mathrm{ff}(t),u_\mathrm{ff}(t),\ulw,\olw)] \\
    & + [\sfM_u^{\xi_\mathrm{nom}(t)}(\ulx,\olx,u_\mathrm{ff}(t),u_\mathrm{ff}(t),\ulw,\olw)]K) (x - x_\mathrm{nom}(t)) \\
    & + [\sfM_w^{\xi_\mathrm{nom}(t)}(\ulx,\olx,u_\mathrm{ff}(t),u_\mathrm{ff}(t),\ulw,\olw)] (w - w_\mathrm{nom}(t)) \\
    &+ f(x_\mathrm{nom}(t),u_\mathrm{ff}(t),w_\mathrm{nom}(t)).
\end{aligned}
\end{align}
Thus, as long as $x_\text{nom}(t)\in[\ulx,\olx]$ and $w_\mathrm{nom}(t)\in[\ulw,\olw]$, 
\begin{align}\label{eq:pendapx:clinclfun}
\begin{aligned}
    [\sfF^\pi&(t,\ulx,\olx,\ulw,\olw)] = ([\sfM_x^{\xi_\mathrm{nom}(t)}(\ulx,\olx,u_\mathrm{ff}(t),u_\mathrm{ff}(t),\ulw,\olw)] \\
    & + [\sfM_u^{\xi_\mathrm{nom}(t)}(\ulx,\olx,u_\mathrm{ff}(t),u_\mathrm{ff}(t),\ulw,\olw)]K) ([\ulx,\olx] - x_\mathrm{nom}(t)) \\
    & + [\sfM_w^{\xi_\mathrm{nom}(t)}(\ulx,\olx,u_\mathrm{ff}(t),u_\mathrm{ff}(t),\ulw,\olw)] ([\ulw,\olw] - w_\mathrm{nom}(t)) \\
    &+ f(x_\mathrm{nom}(t),u_\mathrm{ff}(t),w_\mathrm{nom}(t))
\end{aligned}
\end{align}
is an inclusion function of the closed-loop vector field $f^{\pi}$.

In particular, note that the condition $x_\mathrm{nom}(t)\in[\ulx,\olx]$ is satisfied when~\eqref{eq:pendapx:clinclfun} is used to build an embedding system, as long as the initial condition $x_0\in[\ulx_0,\olx_0]$ used as the initial condition of the embedding system trajectory, and for every $t\in[0,T]$, $w_\mathrm{nom}(t)\in[\ulw,\olw]$ used as the disturbance bounds in the embedding system.

\end{document}